\documentclass[9pt,twoside]{rmaa-rho_light}
\RMxAAtemplatetype{\RMxAA} 

\vol{100}
\pages{30-36}
\thisyear{2025}
\doi{\href{https://doi.org/10.22201/ia.01851101p.20XX.XX.XX.XX}{https://doi.org/10.22201/ia.01851101p.20XX.XX.XX.XX}}

\title{NGC~6426: Colour-Magnitude diagram and variable star population as indicators of metallicity, distance, and age.}

\author[1]{A. Arellano Ferro \orcidlink{0000-0002-8514-2395}}
\author[2]{R. Michel \orcidlink{0000-0003-1263-808X}}
\author[3,4]{D. Deras\orcidlink{0000-0002-3730-9664R}}
\author[5]{M. Vaca Montejano}
\author[6]{S. Muneer\orcidlink{0000-0002-4024-956X}}
\author[7]{I. Bustos Fierro\orcidlink{0000-0001-6032-5446}}

\affil[1]{Instituto de Astronomía, Universidad Nacional Autónoma de México, Ciudad Universitaria, C.P. 04510, Ciudad de México, México}
\affil[2]{Instituto de Astronomía, Universidad Nacional Autónoma de México, AP 106, Ensenada 22800, BC, México}
\affil[3]{Instituto Nacional de Astrof\'isica, \'Optica y Electr\'onica, Luis Enrique Erro No.1, C.P. 72840, Tonantzintla, Pue., M\'exico}
\affil[4]{Secretaría de Ciencia, Humanidades, Tecnolog\'ia e Inovaci\'on, Av. de los Insurgentes Sur 1582, C.P. 03940, Ciudad de México, M\'exico}
\affil[5]{Facultad de Ciencias, Universidad Aut\'onoma de Baja California, Ensenada 22860, C, México}
\affil[6]{Indian Institute of Astrophysics, 2nd block Koramangala, Bangalore, 560034, India}
\affil[7]{ Observatorio Astron\'omico, Universidad Nacional de C\'ordoba, X5000IND C\'ordoba, Argentina}

\leadauthor{Arellano Ferro et al.}
\smalltitle{LaTex Macro RMxAA}

\corres{A. Arellano Ferro}
\email{armando@astro.unam.mx}

\received{April 16th, 2026}
\accepted{\today}

\license{Texto de la licencia aquí}

\setbool{rho-abstract}{true} 
\setbool{rho-resumen}{true} 

\begin{abstract}
   We present new $VI$ CCD photometry of the globular cluster NGC~6426. The membership status of all point sources and, in particular, of all the variables in the field of view is discussed, including four recently detected variables. The double mode nature of variables V3 and V9 is confirmed and their periods are calculated. The presence of a cluster member anomalous Cepheid (V17) is reported. Employing only likely members we produce a decontaminated colour-magnitude diagram (CMD) differentially dereddened, which enables us to compare it with theoretical predictions and to study the variable stars distribution. The mean cluster metallicity and distance were calculated via the RR Lyrae $V$-band light curve Fourier decomposition and an $I$-band calibration of the Period-Magnitude-Metallicity relation. Further, a comparison using a Bayesian approach between theoretical isochrones of  Dartmouth Stellar Evolution Database (DSED),  with a deep HST CMD, gave us independent estimates of distance modulus, metallicity, colour excess, and cluster age.
   
\end{abstract}

\keywords{globular clusters: general, globular clusters: individual: NGC~6426, stars: horizontal branch, stars: fundamental parameters, stars: variables:
RR Lyrae}

\begin{resumen}
Presentamos nueva fotometr\'ia CCD, en las bandas $V$ e $I$, del c\'umulo globular NGC~6426. Analizamos la pertenencia al c\'umulo de todas las fuentes puntuales, particularmente, de todas las variables en el campo de visión, incluyendo cuatro variables detectadas recientemente. Confirmamos la naturaleza de doble modo de V3 y V9 y ofrecemos sus periodos. Reportamos a V17 como una Cefeida anómala miembro del cúmulo. Empleando únicamente los miembros probables, generamos un diagrama color-magnitud desenrojecido diferencialmente, que permite estudiar la distribución de las variables y compararlo con las predicciones te\'oricas. La metalicidad y la distancia medias del c\'umulo se calcularon mediante la descomposición de Fourier de las curvas de luz en la banda $V$ de estrellas RR~Lyrae, as\'i como una calibraci\'on en la banda $I$ de la relaci\'on Periodo-Magnitud-Metalicidad. Adem\'as, una comparaci\'on bayesiana de las is\'ocronas te\'oricas de la base de datos de evoluci\'on estelar de Dartmouth (DSED), con un diagrama color-magnitud profundo del HST, nos proporcion\'o estimaciones independientes del m\'odulo de distancia, la metalicidad, el exceso de color y la edad del c\'umulo.

\end{resumen}

\begin{document}

\maketitle
\pagestyle{fancy}\thispagestyle{firststyle}

\section{INTRODUCTION}

NGC~6426 is a distant outer Galactic halo globular cluster (GC) currently located at about 20 kpc from the Sun and 5.8 kpc north of the Galactic plane towards the constellation of Ophiuchus ($l = 28^{\circ}.09$,   $b= 16^{\circ}.23$; $\alpha = 17^{\mbox{\scriptsize h}}
44^{\mbox{\scriptsize m}} 54.65^{\mbox{\scriptsize s}}$, $\delta = +03^{\circ} 10\arcmin
12.5\arcsec$, J2000). Its galactic orbit over the last 500 Myr shows excursions of up to 10-15 kpc from the Galactic centre (Baumgardt et al. 2023)\footnote{%
 \texttt{https://people.smp.uq.edu.au/HolgerBaumgardt/globular/}}. It displays a Colour Magnitude Diagram (CMD) with a Turn Off point at $V \sim 21.5$ mag and a well populated Horizontal Branch (HB) at about $V \sim 18.5$ mag. It is among the oldest GC, with an age of 13.0$\pm$1.5 Gyr \citep{Dotter2011} and one of the most metal poor in the Galaxy with [Fe/H]$\sim -2.15$ \citep{Harris1996}. A detailed analysis of the age-metallicity relation for inner and outer halo GC's led  \citet{Dotter2011} to conclude that a rapid chemical enrichment took place in the inner halo while a prolonged GC formation occurred in the outer halo. These authors argue that this picture is consistent with the outer halo GCs being formed in dwarf galaxies and later being accreted by the Milky Way. Comparisons of observed GC distribution with hydrodynamical
simulations of GCs formation and evolution, the splitting of the distribution in the energy-angular momentum space, and age-energy considerations allowed several authors to associate families of GCs with identified accretion events. We refer to the work of \citet{Callingham2022}, where NGC~6426 has been associated to the Helmi Stream \citep{Helmi1999,Helmi2018}.
 
The instability strip, at the crossing with the HB is populated by a well identified group of RR Lyrae stars. The Catalog of Variable Stars in Globular clusters (GVSGC, \citet{Clement2001}) lists 10 RRab, 5 RRc, and one RRd stars, most of them apparently being likely cluster members. Our aims in the present work include to use our CCD $VI$ photometry to investigate the CMD structure and to estimate the average metallicity, distance, and age of the cluster. To achieve a proper unreddened CMD free of field stars, a membership analysis is performed based on $Gaia$-DR3 positions and proper motions. The effects of differential reddening are estimated. The metallicity and distance of cluster member RR Lyrae stars are anchored, via the Fourier decomposition of their light curves, to the scales
discussed by \citet{Arellano2024}  for a family of 39 globular
clusters. We highlight the membership status and the classification of all variable stars measured in our data and address the distribution of RRab and RRc stars on the HB in connection with the Oosterhoff type of the cluster.  A Bayesian
approach to the comparison of theoretical isochrones with a deep Hubble Space Telescope (HST) CMD, enables 
independent estimates of distance modulus, metallicity, colour excess, and cluster age. The paper is structured as follows: Section~\S~\ref{sec:observations} describes the observations and datasets used in this work. Section~\S~\ref{sec:reductions} details the image processing procedures applied to our datasets and the transformation to their corresponding standard system. Section~\S~\ref{sec:var_stars} presents the population of variable stars in NGC~6426 and the determination of their physical parameters via Fourier decomposition of their light curves. Section~\S~\ref{sec:CMD} presents the CMDs of NGC~6426 in the $VI$ and ACS photometric systems, and discusses their stellar membership, the cluster's reddening, and the determination of its age via isochrone fitting. Finally, Section~\S~\ref{sec:conclusions} summarises our results and conclusions.

\section{OBSERVATIONS}
\label{sec:observations}
The observations were conducted with the 0.84m telescope of San Pedro M\'artir Observatory, Baja California, Mexico (SPM), during 20  nights divided in five seasons between 2012 and 2025. 
The telescope was equipped with a \textit{Marconi 5} CCD detector (an e2v CCD231-42 chip with $15\times15$ $\mu^ 2$ pixels, gain of 2.2e$^-$/ADU and readout noise of 3.6e$^-$). The field of view was $9^{\prime}\times9^{\prime}$ and binning 2$\times$2 was employed during all the observations. Flats field and bias frames were also taken during each observing night.

Another set of observations was obtained at the Indian Astronomical Observatory (IAO; Hanle, India) on 2024 June 3rd and 8th,  and 2025 May 20th, using the 2\,m telescope equipped with an E2V CCD44-82-0-E93 detector ($2048 \times 4096$\,px; $0\farcs296$\,px$^{-1}$). The field of view of approximately $10\farcm1\times10\farcm1$. 

A total of 961 and 953 images were secured in the $V$- and $I$-bands, respectively. The detailed log of observations is given in Table~\ref{tab:log}.

We also made use of an ancillary set of images from the Advanced Camera Survey (ACS) obtained by the HST in the $F606W$ and $F814W$ filters (GO:11586; PI:Dotter). This data set is comprised by one 45s and four 500s images in the $F606W$ filter, and by one 50s and four 540s images in the $F814W$ filter.

\begin{table}[t]
\begin{center}
\caption{Log of observations of NGC~6426. Typical exposure times in SPM were 180s and 120s in the $V$ and $I$ filters, respectively while they were 480s and 180s in IAO. We list the average nightly seeing in column 4.}
\label{tab:log}
    \begin{tabular}{ccccc} 
    \hline
    Date & N$_V$ &  N$_I$ & Avg. seeing ($^{\prime \prime}$)& Site \\
    \hline
    15/08/2012 & 49 & 49 & 2.1& SPM \\
    16/08/2013 & 45 & 45 & 2.0& SPM \\
    17/08/2013 & 18 & 18 & 2.2& SPM \\
    17/07/2024 & 51 & 52 & 2.0& SPM \\
    18/07/2024 & 36 & 36 & 1.8& SPM \\
    19/07/2024 & 47 & 48 & 1.9& SPM \\
    06/05/2025 & 16 & 97 & 1.8& SPM \\
    07/05/2025 & 54 & 54 & 1.9& SPM \\    
    08/05/2025 & 57 & 58 & 1.8& SPM \\    
    09/05/2025 & 58 & 58 & 2.1& SPM \\
    10/05/2025 & 39 & 39 & 2.4& SPM \\
    11/05/2025 & 31 & 32 & 1.8& SPM \\
    15/05/2025 & 50 & 50 & 1.8& SPM\\
    16/05/2025 & 48 & 48 & 1.7& SPM \\
    02/06/2025 & 48 & 47 & 1.8& SPM \\
    03/06/2025 & 63 & 60 & 1.7& SPM \\
    05/06/2025 & 37 & 38 & 1.6& SPM \\
    06/06/2025 & 59 & 61 & 2.0& SPM \\
    24/06/2025 & 69 & 69 & 2.0& SPM \\
    25/06/2025 & 58 & 60 & 2.7& SPM \\
    03/06/2024 & 5 & 4 & 1.9& IAO \\
    08/06/2024 & 18 & 20 & 2.1& IAO \\
    20/05/2025 & 6 & 4 & 2.3& IAO \\
    \hline
    Total: & 961 & 953 &  \\
    \hline
\end{tabular}
\end{center}
\end{table}

\section{REDUCTIONS AND TRANSFORMATION TO THE STANDARD SYSTEM}
\label{sec:reductions}

The SPM $VI$ images were processed with the help of the {\tt \string IRAF}\footnote{{\tt \string IRAF} (\url{http://iraf.noao.edu/}) is distributed by the National Optical Astronomy Observatories, which are operated by the Association of Universities for Research in Astronomy, Inc., under cooperative agreement with the National Science Foundation.} package. After correcting each image for bias and flat field, images were all aligned and a reference frame for each filter was produced by averaging all of them. The stars' XY coordinates were determined in the reference frame and then instrumental magnitudes were calculated using the standard PSF photometry method. Instrumental magnitudes were then transformed to apparent magnitudes through differential photometry relative to either of the two reference stars: 2MASSJ17445132+0310444 ($V$=14.513, $I$=13.366) for stars with $V-I < 1.3$ and 2MASSJ17445160+0310057 ($V$=15.325, $I$=13.624) for the redder stars. The apparent magnitudes of the reference stars and the $V-I$ colours of the field stars were obtained from the Gaia synthetic photometry catalogue \citep{GaiaCollaboration2023Synthetic}. 

The photometric analysis of the ACS data set was carried out on the -flc images,which are corrected for dark-subtraction, flat-field, bias, and charge transfer efficiency. The standard package \textrm{DAOPHOT IV} \citep{Stetson1987} was used for the data reduction. We start by modeling the shape of the Point Spread Function (PSF) by selecting 200 isolated stars present in our images. For the ACS, its Full Width at Half Maximum (FWHM) is set at 2 pixels ($\sim0\arcsec.1$), and we sampled each of these selected stars within a 20-pixel radius (equivalent to $\sim1\arcsec$) for the ACS. The best PSF model for each image (based on a $\chi^2$ statistic) was provided by a Moffat function \citep{Moffat1969} for the $F606W$ filter and a Penny function \citep{Penny1976} for the $F814W$ filter.

These models were applied to all the sources detected in the images of the first data set above a 5$\sigma$ threshold from the local background level. We subsequently created a master catalogue, which included the instrumental magnitudes and positions of each stellar source. Using DAOPHOT/ALLFRAME \citep{Stetson1994}, a fit was enforced in each image at the corresponding positions of all these sources. The results were then combined using DAOPHOT/DAOMASTER to ultimately achieve homogenised instrumental magnitudes and their associated photometric errors.

The instrumental positions of the stars present in the ACS data set were corrected for geometric distortions in both chips, following the procedure described by \citet{Anderson2006}. We then assigned celestial coordinates (RA and Dec) to the new instrumental positions through cross-correlation with the stars in common with the $Gaia$-DR3 catalogue \citep{Gaia2022}.

The instrumental magnitudes of the ACS data sets have been calibrated to the VEGAMAG system using the date-dependent\footnote{\url{https://acszeropoints.stsci.edu/}} photometric zero point values: ZP$_{\rm F606W1,2}$ = 26.407, ZP$_{\rm F814W1,2}$ = 25.510. Finally, we applied independent aperture and encircled energy corrections for each chip and filter.


\section{THE VARIABLE STARS}
\label{sec:var_stars}
\begin{figure*}[ht]
\begin{center}
\includegraphics[width=16cm]{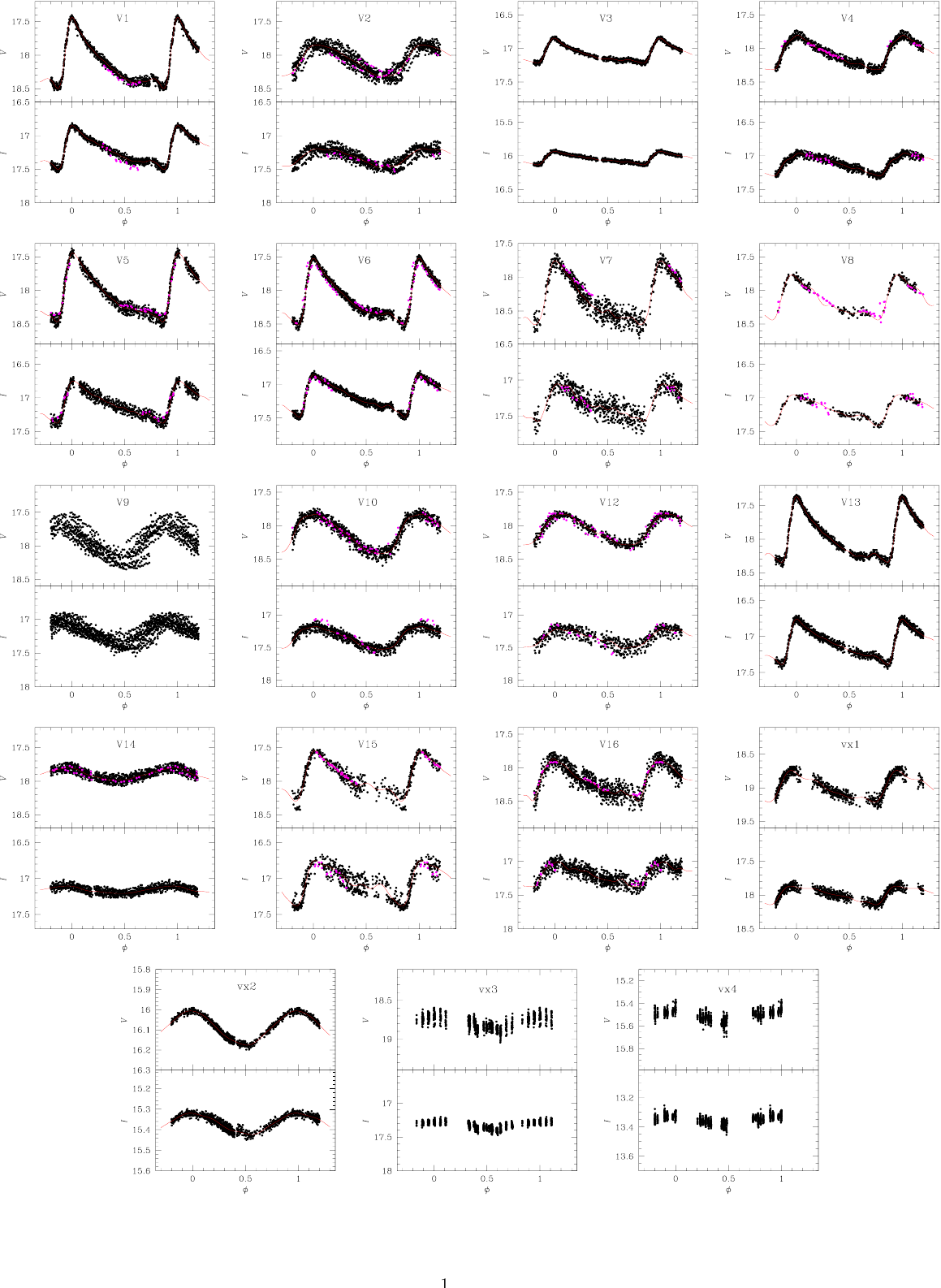}
\caption{Light curves of variable stars in NGC~6426. Black and magenta dots correspond to SPM and Hanle data,  respectively. The continuous red lines are the Fourier fits according to Eq. \ref{eq:Foufit} and their coefficients are listed in Table~\ref{tab:fourier_coeffs}. With the exception of vx2 and vx4, the vertical scale is the same for all stars in order to note their amplitude differences.}
\label{Fig:Vars}
\end{center}
\end{figure*}

\begin{figure*}[ht]
\begin{center}
\includegraphics[width=12cm]{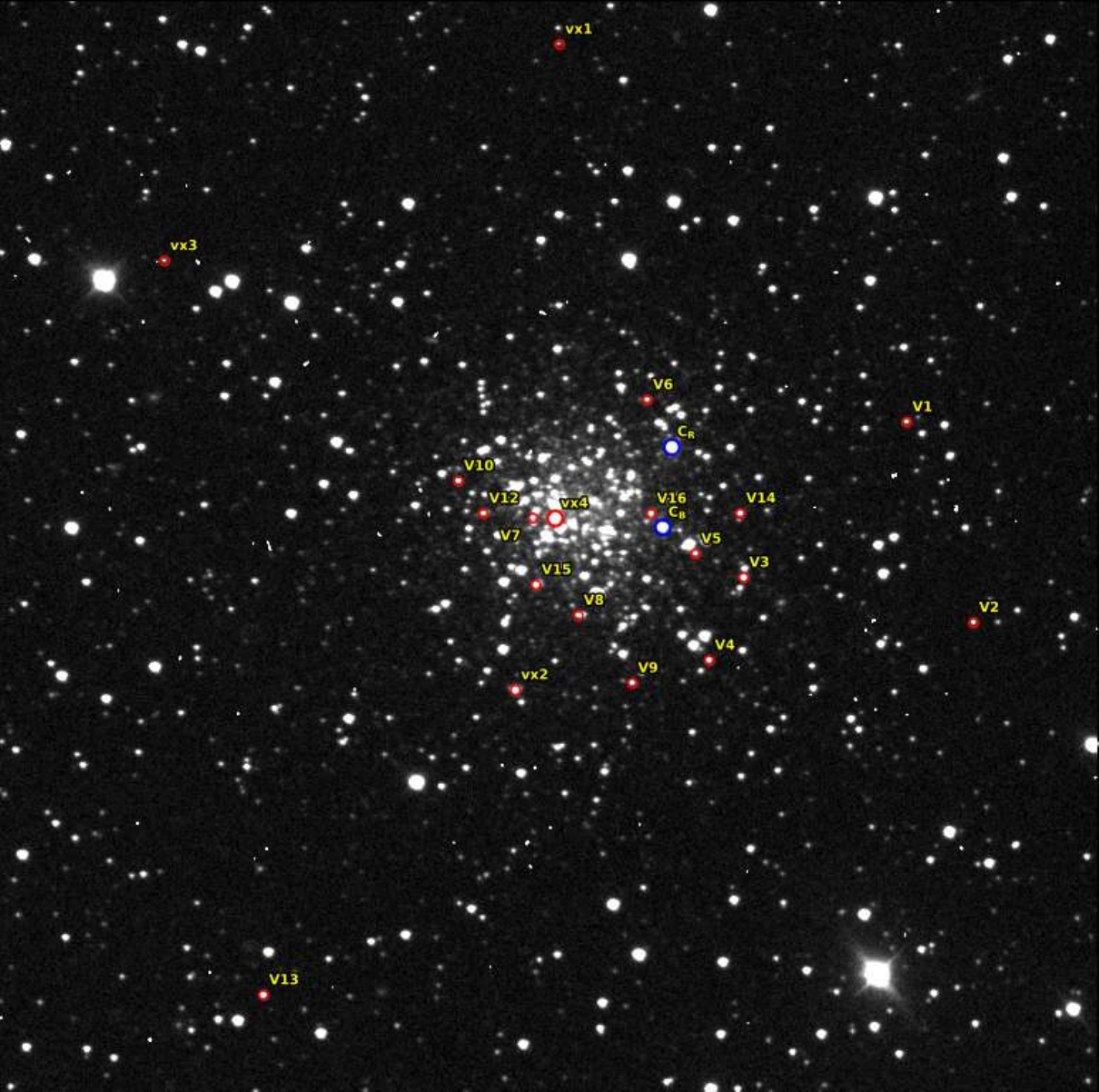}
\caption{Identification chart of variable stars in NGC~6426. The star V11 is not in the field of our images.  The two stars labeled C$_B$ and  C$_R$ are the blue and red stars used for the differential photometry described in \S~\ref{sec:reductions}. In the image, North is up and East is to the left.}
\label{Fig:IDchart}
\end{center}
\end{figure*}

The light curves of the variable stars in the field NGC~6426 are shown in Fig.~\ref{Fig:Vars}. Variables V1 to V16 are reported in the CVSGC. We note that the variable V11 is out of the FoV of our images and therefore is not studied in the present work. 
We detected the variability in four stars not included in the  2015 edition of the CVSGC and labeled them as vx1, vx2, vx3, vx4. Employing the coordinates determined on our calibrated reference images we confirmed that all four new detections have in fact been reported in previous works. We shall defer detail discussion of these four stars to the Appendix~\ref{ap:appendix}. Particular attention is called to the classification of vx2, which appears to be a cluster member anomalous Cepheid. In Table~\ref{tab:datosgenerales1} the mean intensity weighted magnitudes, peak to peak amplitudes, ephemerides, sky coordinates, and membership status are reported for each variable. In Fig.~\ref{Fig:IDchart} we offer an identification chart of all the above mentioned variables. 

\begin{table*}
    \begin{center}
    \caption{Variable stars in the Field of NGC~6426. The membership flags in column 11 are m=likely cluster member; f=likely field star} 
    \begin{tabular}{ccccccccccc}
    \hline
    Variable & Type & $\langle$\textit{V}$\rangle$ & $\langle$\textit{I}$\rangle$ & $A_{V}$ & $A_{I}$ & Period & $HJD_{max}$ &RA & Dec. &Memb.  \\
     &   & (mag) & (mag) & (mag) & (mag) & (days) & (246 0000.+) &(J2000.0) & (J2000.0) & (m/f)\\
    \hline
\hline
V1	& RRab	 &	18.128	& 17.226	& 1.064	& 0.671	& 0.617837 & 0850.9274 & 17:44:43.76& +03:10:56.79& m \\
V2	& RRd &	18.086&17.326 &0.508 &0.288 &0.355447 & 0806.8779&17:44:41.59&+03:09:20.29 & m \\
V3	& RRab &17.086& 16.042&0.374& 0.200&0.678809 &0904.7056 &17:44:48.99&+03:09:41.63 & f \\
V4	& RRab &18.091&17.117 &0.547 & 0.352&0.768852 &0850.7027  &17:44:50.09&+03:09:01.69 & m \\
V5	& RRab &18.062&17.092 &1.063 & 0.662&0.709066 &0803.8912   &17:44:50.54&+03:09:53.00 & m \\
V6 & RRab &	18.129&17.185 &1.033 & 0.644&0.681989 &0850.7833   &17:44:52.11&+03:11:07.01 & m \\
V7	& RRab &18.340&17.335 &0.952 &0.618 &0.753658 & 0510.6818   &17:44:55.77&+03:10:09.95 & m \\
V8	& RRab &18.166&17.177 &0.662 & 0.474&0.706260& 0810.9054  &17:44:54.31&+03:09:23.09 & m \\
V9	& RRd  &17.956&17.205 &0.615 &0.392&0.289673 &0805.9644  &17:44:52.56&+03:08:50.63 & m \\
V10	& RRc &	18.131&17.355 &0.630 &0.387 &0.364980 &0510.7595 &17:44:58.18&+03:10:27.66 & m \\
V12 & RRc &18.073&17.358 &0.486 & 0.337&0.310512 &0832.7723 &17:44:57.37&+03:10:12.03 & m \\
V13	& RRab &18.003&17.112 &0.930 &0.692 &0.651908 &0903.8003  &17:45:04.39&+03:06:19.80 & m? \\
V14	& RRc &17.903& 17.162&0.210 &0.130 &0.291760 &0851.9148   &17:44:49.10&+03:10:12.54 & m \\
V15	& RRab &17.962&17.051 &0.725 &0.533&0.689934& 0903.7791  &17:44:55.67&+03:09:37.75 & m \\
V16	& RRab &18.208&17.204 &0.676 &0.404&0.748060 &0902.8398   &17:44:51.97&+03:10:12.22 & m \\
vx1	& RRab &19.008&17.994 &0.252 &0.314 &0.665117& 0828.9623   &17:44:54.98&+03:13:58.15 & f \\
vx2(V17)& AC &16.090&15.368 &0.173 &0.114 &1.073675 &0806.8295   &17:44:56.32&+03:08:47.15 & m \\
vx3	& -- &	18.783&17.327  &0.265 &0.163 &18.084& 0804.8437   &17:45:07.66&+03:12:13.19 & f \\
vx4(V18)	& SR &	15.505&13.350  &0.114 &0.064 &39.21& 0832.9548  &17:44:55.06 &+03:10:09.85 & m \\

\hline
    \hline
    \end{tabular}
    \label{tab:datosgenerales1}
    \end{center}
\end{table*}

\subsection{RR Lyrae Fourier decomposition and physical parameters}

In past years, aiming to determine the metallicity and distance of GC's in homogeneous scales from CCD photometry, we have employed the Fourier decomposition of the light curves of cluster member RR Lyrae stars in nearly 40 clusters. The Fourier parameters were used along $ad~hoc$ semiempirical calibrations to estimate individual [Fe/H] and $M_V$, masses, and radii. The specific calibrations for [Fe/H] come from the works of \citet{Jurcsik1996} for RRab stars and \citet{Morgan2007} for RRc stars. The calibrations for $M_V$ come from \citet{Kovacs2001} and 
\citet{Kovacs1998} for the RRab and RRc stars,  respectively.
The interested reader can find our results for a family of globular clusters and a detailed description of the semi-empirical calibrations employed and their zero points for RRab and RRc stars, in the works of \citet{Arellano2010} and \citet{Arellano2024}. It has been shown by \citet{Arellano2024} that the calibrations and zero points cited above
 produce, within the respective uncertainties, iron abundances concordant with the UVES spectroscopic scale \citep{Carretta2009} and distances matching those of \citet{Baumgardt2021}. 

The Fourier series for the light curves decomposition has the form

\begin{equation}
\label{eq:Foufit}
m(t) = A_0 + \sum_{k=1}^{N}{A_k \cos\ [{2\pi \over P}~k~(t-E) + \phi_k] },
\end{equation}

\noindent
where $m(t)$ is the magnitude at time $t$, $P$ is the period, and $E$ is the epoch. A
linear
minimization routine is used to derive the best-fit values of the 
amplitudes $A_k$ and phases $\phi_k$ of the sinusoidal components. 
Then, the amplitudes and phases of the harmonics in Eq.~\ref{eq:Foufit} are turned into the 
Fourier parameters, defined as $\phi_{ij} = j\phi_{i} - i\phi_{j}$, and $R_{ij} =
A_{i}/A_{j}$. The Fourier fits are shown as a red continuous line in the curves of Fig.~\ref{Fig:Vars}, and the relevant Fourier coefficients are listed in Table~\ref{tab:fourier_coeffs}. Also listed in Table~\ref{tab:fourier_coeffs} is the $deviation parameter ~D_m$. If $D_m < 3.0$ then the light curve under analysis is considered to satisfy the compatibility condition with the [Fe/H] calibration for RRab stars \citep{Jurcsik1996, Kovacs_Kanbur1998}. In order to mantain a more solid sample we slightly relaxed the criterion to  $D_m < 5.0$ Then, the application of the calibrations and zero points described in the works of \citet{Arellano2010,Arellano2024}, lead to the resulting physical parameters reported in Table~\ref{tab:fisicos}.

\begin{table*}
\caption{Fourier coefficients of the $V$ light curves of RR Lyrae stars in Fig.~\ref{Fig:Vars}. Numbers between parenthesis refer to errors to the last decimal corresponding places. For the description of the $D_m$ deviation parameter refer to \S 4.1}
\centering                   
\begin{tabular}{lllllllllr}
\hline
Variable ID     & $A_{0}$    & $A_{1}$   & $A_{2}$   & $A_{3}$   & $A_{4}$   &
$\phi_{21}$ & $\phi_{31}$ & $\phi_{41}$ 
&  $D_{\mbox{\scriptsize m}}$ \\
     & ($V$ mag)  & ($V$ mag)  &  ($V$ mag) & ($V$ mag)& ($V$ mag) & & & & \\
\hline
       &       &   &   &   & RRab star    & &        
   &             &       \\
\hline
V1 & 18.128(1)& 0.365(2)& 0.159(2)& 0.128(2)& 0.087(2)& 3.853(14)& 7.941(19)& 5.893(26) & 0.8\\
V4&  18.091(1)& 0.208(2)& 0.086(2)& 0.042(2)& 0.015(2)& 4.324(28)& 8.851(51)& 7.766(121)& 4.9\\
V5&  18.062(2)& 0.362(3)& 0.179(3)& 0.118(3)& 0.072(3)& 3.958(21)& 8.333(31)& 6.341(48)& 2.5\\
V6&18.129(1) &0.336(2)& 0.168(2)& 0.119(2)& 0.078(2)& 3.872(16)& 8.269 (23)& 6.295(33)&1.5\\
V7&18.340(4)& 0.360(6)&0.177(6)& 0.085(6)& 0.064(6)& 3.874(47)& 8.148(83)& 6.246(108)&4.3\\
V8&18.166(5)& 0.253(34)&0.121(83)& 0.064(65)& 0.051(34)& 3.821(99)& 8.216(99)&  6.649(99)&10.0$^1$\\
V13&18.003(1)&  0.343(2)&  0.156(2)&  0.118(2)& 0.083(2)&  3.913(15)&  8.041(21)&  6.077(28)&  1.3\\

V16&18.208(3)& 0.230(4)&0.100(4)& 0.069(4)& 0.027(4)& 4.134(58)& 8.409  (86)& 7.386(178)&3.2\\

vx1&19.008(4)& 0.189(5)&0.065(4)& 0.048(4)& 0.033(4)& 4.199(96)& 8.857(96)& 6.879(149)&--\\
\hline
             &                    &           &           & &RRc stars            
 &             &             &\\
\hline
V2& 18.086(2)& 0.231(4)& 0.049(4)& 0.017(4)& 0.014(4)& 4.771(85)&2.699(219)& 1.213(287)\\
V10& 18.131(2)& 0.290(3)& 0.063(3)& 0.026(3)& 0.016(3)&4.667(48)&2.861(110)& 1.832(180)\\
V12& 18.072(2)& 0.220(3)& 0.055(3)& 0.008(3)& 0.007(3)& 4.425(54)& 2.632(322)& 1.284(362)\\

V14&17.903(2)& 0.087(2)& 0.010(2)& 0.006(2)& 0.002(2)&  2.523(219)& 4.917(358)& 0.926(900) \\

\hline	
\hline
\end{tabular}
\label{tab:fourier_coeffs}

\raggedright
\center{
1. Not included in physical parameters calculation.\\
}
\end{table*}

\begin{table*}
\begin{center}
\caption[] {\small Physical parameters of the RRab and RRc stars from the Fourier decomposition of their light curves. Only stars that were identified as cluster members are considered in these calculations $^*$}
\label{tab:fisicos}
\hspace{0.01cm}
 \begin{tabular}{lllllllll}
\hline 
 &  &  & RRab star &  & & & \\
\hline
Star&[Fe/H]$_{\rm ZW}$&[Fe/H]$_{\rm UVES}$ & $M_V$ & log~$T_{\rm eff}$  & log$(L/{L_{\odot}})$ & D (kpc)&$M/{M_{\odot}}$&$R/{R_{\odot}}$\\

\hline
V1 &-1.95(3) &-2.02(4)& 0.485(3)& 3.800(7) & 1.731(1) & 20.98(3)&0.80(7)& 5.72(1)\\
V4 &-1.67(6) &-1.62(7)& 0.418(3)& 3.784(15) & 1.748(1) & 21.32(3)&0.74(13)& 6.50(3)\\
V5 &-1.93(4) &-1.99(5)& 0.368(4)& 3.795(9) & 1.776(2) &21.43(5) &0.80(8)& 6.20(2)\\
V6 &-1.89(3) &-1.93(4)& 0.431(3)& 3.796(8) & 1.750(1) & 20.98(3)&0.77(7)&6.62(1)\\
V7$^1$ &-2.27(8) &-2.55(13)& 0.293(9)& 3.786(14) & 1.820(3) & 24.72(11)&0.92(15)& 6.42(4)\\
V8$^1$ &-2.03(94) &-2.14(94)& 0.453(66)& 3.787(114) & 1.746(27) &21.10(65) & 0.80(8)& 6.19(47)\\
V13 &-1.99(3) &-2.08(4)& 0.459(3)& 3.796(8) & 1.742(1) & 20.33(3)& 0.80(7)& 5.91(1)\\
V16 &-2.00(8) &-2.10(12)& 0.437(6)& 3.780(21) & 1.752(2) & 21.44(6)& 0.81(20)& 6.40(4)\\

\hline
Weighted Mean&-1.93&-1.97&0.440&3.796& 1.746& 20.96& 0.79 & 6.47\\

$\sigma$ &$\pm$ 0.13  &$\pm$ 0.13 & $\pm$ 0.040 &$\pm$ 0.009 &$\pm$ 0.001& 0.44&$\pm$ 0.03 & $\pm$ 0.04 \\
\hline

 &  &  & RRc stars &  & & & \\
 \hline
Star&[Fe/H]$_{\rm ZW}$&[Fe/H]$_{\rm UVES}$ & $M_V$ & log~$T_{\rm eff}$  & log$(L/{L_{\odot}})$ & D(kpc) &$M/{ M_{\odot}}$&$R/{R_{\odot}}$\\

\hline
V2 &-2.06(42) &-2.19(62) &0.519(18) &3.855(1) &1.693(7) & 20.60(17)&0.55(1)  &4.40(4)   \\
V10 &-2.07(22) &-2.20(32) &0.504(14) &3.854(1) &1.698(5) & 20.41(13)&0.54(1)  &4.64(3)  \\
V12  &-1.80(57) &-1.79(73) &0.605(14) &3.863(2) &1.658(5) &19.11(12) &0.55(1) &4.26(3)  \\
V14$^1$ &0.09(73) &-0.40(10) &0.730(10) & 3.880(1)&1.608(5) &16.67(10) &0.43(1)& 3.81(19) \\

\hline
Weighted Mean&-2.04&-2.16&0.546&3.855& 1.681&19.85 &0.54 & 4.47\\
$\sigma$ &$\pm$ 0.17  &$\pm$ 0.17 & $\pm$ 0.050 &$\pm$ 0.001 &$\pm$ 0.003&0.70& $\pm$ 0.01 & $\pm$ 0.01 \\
\hline
\end{tabular}
\end{center}
\raggedright
\center{
1. Not included in the averages. *Numbers in parenthesis indicate the uncertainty in the last decimal place\\
}
\end{table*}

\subsection{ Cluster distance from I-band PMZ calibration}

We can perform yet another independent estimation of the cluster distance via the recent Period-Magnitude-Metallicity (PMZ) calibration of \citet{Prudil_Kunder2024}. Their equation 19 is:

\begin{equation}
	\label{eq:P24}
	M_I = -1.292~log_{10} P +0.196 {\rm [Fe/H]} +0.197.
\end{equation}

We adopted $R_{VI}=1.205$ from \citet{Prudil2025}, from which a distance 

\begin{equation}
	\label{eq:Dist}
	d_I=10^{(1+ 0.2 (I_0-M_I-R_{VI}*E(V-I)))}.
\end{equation}

\noindent
can be calculated. These equations have been applied to the 6 RRab and 3 RRc stars listed in Table~\ref{tab:fisicos}, that were included in the physical parameters averages. For $E(B-V)$ we adopted the mean value 0.34, plus the calculated differential reddening (see \S~\ref{sec:DifRed}). The mean resulting distance is: 21.78 $\pm$ 0.74 kpc

\subsection{The Oosterhoff type}
\begin{figure}[h]
\begin{center}
\includegraphics[width=8cm,height=11cm]{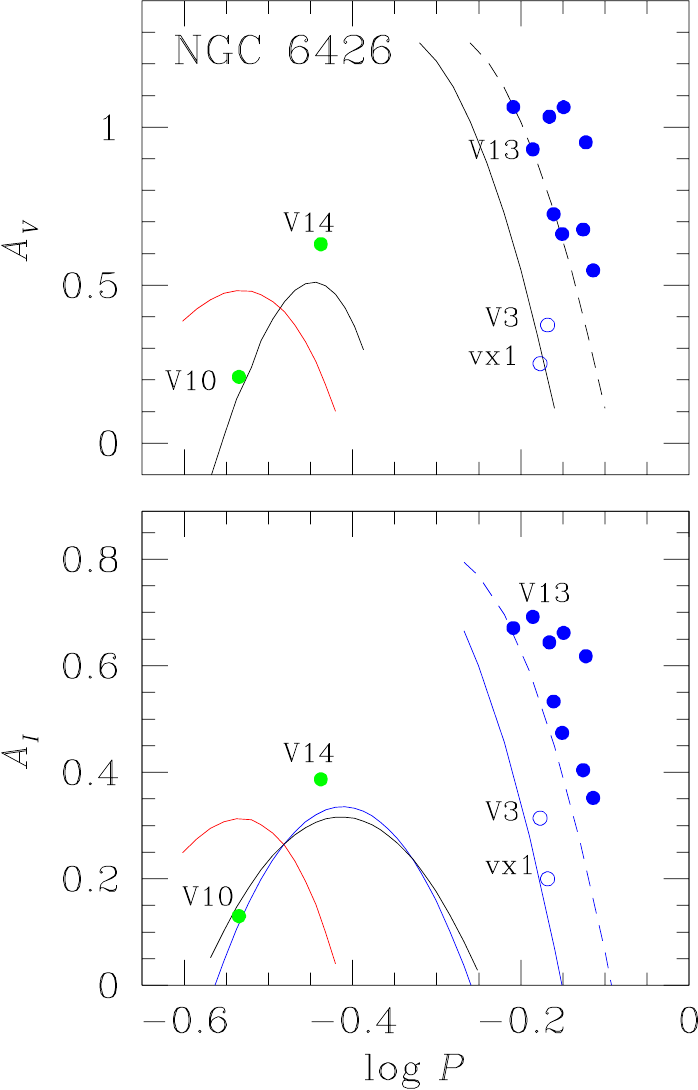}
\caption{Period-Amplitude diagram of NGC~6426. Blue and green dots are used for RRab and RRc stars. Non-member stars are displayed with empty dots. In the top panel the continuous and dashed lines are the loci for unevolved and evolved stars respectively in M3 \citep{Cacciari2005}. The black and red parabolas were calculated for RRc stars in OoII \citep{Kunder2013b} and OoI clusters \citep{Arellano2015} respectively. Similarly, in the bottom panel shows the loci for unevolved and evolved RRab stars \citep{Kunder2013a} and the blue \citep{Yepez2020} and black \citep{Deras2019} parabolas for RRc stars in OoII clusters.}
\label{Fig:Bailey}
\end{center}
\end{figure}

The OoII-type nature of NGC~6426 has been established for a long time (e.g., by \citealt{Papadakis2000}). This classification is well supported by several indicators: first, the average period of the nine member RRab stars listed in Table~\ref {tab:datosgenerales1}, $0.703 \pm0.046$ d; second, the Period - Amplitude diagram (Fig.~\ref{Fig:Bailey}) shows the RRab and RRc stars distributed along the evolved sequences typical of OoII type clusters, with the two modes RRab and RRc stars neatly segregated by the First Overtone Red Edge (FORE) of the instability strip \citep{Arellano2018, Yepez2020}; and third, its iron abundance [Fe/H] $\leq -1.5$ dex. Thus we confirm that these criteria are clear and sufficient to assign the OoII type to NGC~6426.

\section{THE COLOUR-MAGNITUDE DIAGRAM OF NGC~6426}
\label{sec:CMD}
\subsection{Stellar membership}
\label{sec:Membership}

\begin{figure*}[ht]
\begin{center}
\includegraphics[width=16cm,height=6.5cm]{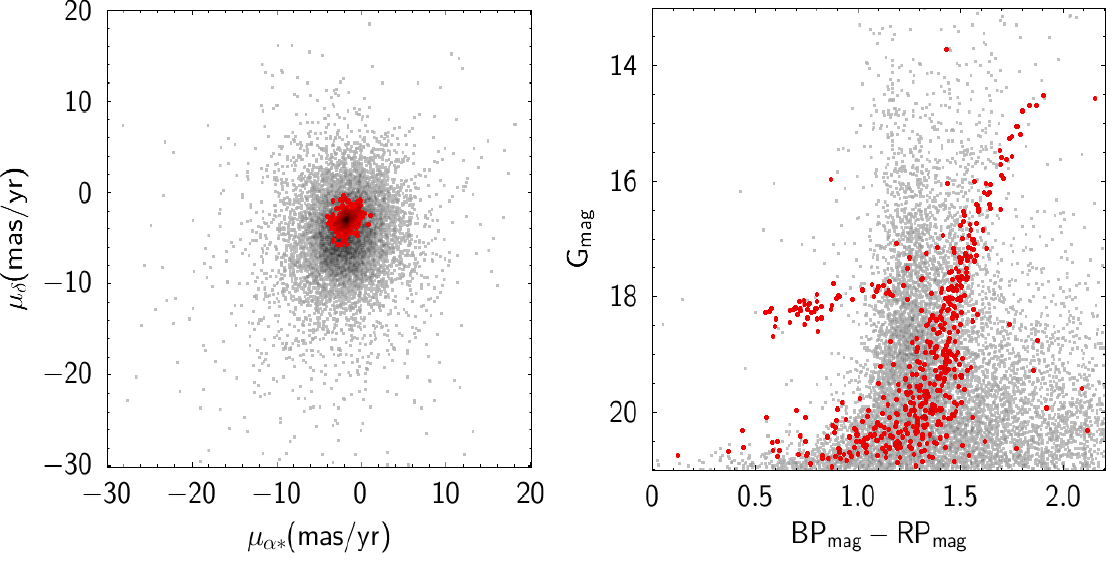}
\caption{VPD and CMD diagrams of NGC~6426 constructed using $Gaia$-DR3 proper motions and magnitudes (see \S~\ref{sec:Membership}). Grey and red dots represent field and likely member stars, respectively.}
\label{Fig:membership}
\end{center}
\end{figure*}

In order to produce a CMD free of field stars, it is possible to employ the proper motions available in the $Gaia$-DR3 catalogue. In the present work, we followed the procedure of \citet{Bustos2019} to separate the likely members from the field stars. We included stars within a radius of 15$\arcmin$ from the cluster centre and found 10,152 stars with proper motions available, 587 of which were identified as possible members. These members are shown with red dots on the VPD and CMD diagrams of  Fig.~\ref{Fig:membership} where a clear red giant branch (RGB) and a well developed HB are depicted.

From the above analysis we also conclude the membership status of 16 known variables in the CVSGC, and list them in Table~\ref{tab:datosgenerales1}. These membership results are independent of the membership flags shown in the CVSGC which where assigned by \citet{Prudil2024} after a critical evaluation for variables in globular clusters from the bulky membership analysis by \citet{Vasiliev2021}. Given that the membership determination approaches of \citet{Bustos2019}, employed in the present paper, and the one followed by \citet{Vasiliev2021} are different, the results are independent. In the case of these 16 variables in NGC~6426, we see that the results match well.

\subsection{Differential Reddening}
\label{sec:DifRed}

The CMD of NGC~6426 shows signs of differential reddening \citep{Dotter2011}. The presence of differential reddening may be a limiting factor in estimating the age of the cluster by means of isochrone fitting, as it broadens and distorts the main features of its CMD. Therefore, we attempted to take this into account before estimating the age of the cluster and other parameters using the different approaches considered here. Estimates of  the average overall reddening can be obtained from the calibrations of \citep{Schlafly2011} $E(B-V)\sim 0.30$ and \citep{Schlegel1998}  $E(B-V)\sim 0.35$. We constructed two differential reddening maps, one based on our ground-based $VI$ photometry and another based on the HST/ACS observations. It is important to note, however, that the sizes of the two photometric catalogues are substantially different: the $VI$ catalogue contains approximately 350 stars, whereas the ACS catalogue contains 19,584 stars. Although a differential reddening map could be constructed for each dataset, the much smaller number of stars in the $VI$ catalogue results in a significantly lower spatial resolution compared with the ACS-based map. For the $VI$ photometric catalogue, we proceeded as follows: First, we derreddened the CMD made of only the member stars by adopting a mean $E(B-V)$ = 0.34. Then, we performed a rough selection of stars along the Main Sequence, Sub Giant Branch and Red Giant Branch, i.e., between $14.0~<~V_{0}~<~19.5$. This selection was divided vertically into bins of 0.5 mag, for which we estimated the 3$\sigma$-clipped medians of $V_{0}$ and $(V-I)_{0}$. These medians were then interpolated to create a Main Ridge Line (MRL), which was used as a reference to determine the geometric distance $\Delta X$ to each cluster member, along the direction of the reddening vector. The reddening vector is defined using the extinction coefficients $R_{V}$ = 3.1 and  $R_{I}$ = 1.8 \citep{Cardelli1989}. For each cluster member, we identified the $n$ nearest reference stars and computed the 3$\sigma$-clipped median of their geometric distances to the MRL, along the reddening vector. This median value corresponds to the star’s assigned displacement $\Delta X$, which
is transformed into the relative differential reddening $\delta E(B-V)$
by using:

\begin{equation}
    \delta E(B-V) = \frac{\Delta X}{\sqrt{R^{2}_{V} + (R_{V}-R_{I})^{2}}}
\end{equation}

To increase the spatial resolution we iteratively performed this computation two times using the $n$ = 50 and 25 closest stars to each cluster member. The differential variations of the colour excess within the sampled field of view are significant, ranging between $-0.03<\delta E(B-V)<0.04$. 
For the ACS photometric catalogue, only a differential reddening correction was applied to the CMD. The MRL was constructed using stars within the magnitude range of $17.0~<~m_{F606W}~<~25.0$, with the following equation:

\begin{equation}
    \delta E(B-V) = \frac{\Delta X}{\sqrt{R^{2}_{F606W} + (R_{F606W}-R_{F814W})^{2}}},
\end{equation}

\noindent
using the extinction coefficients $R_{F606W}$ = 2.8 and  $R_{F814W}$ = 1.8 \citep{Cardelli1989}.
In this case, we iteratively performed this computation three times using the $n$ = 100, 50 and 25 nearest reference stars. An absolute dereddening correction was not performed because this CMD is intended exclusively for isochrone fitting in the following section.

\subsection{Isochrone fitting}
The correction for differential reddening in the CMD of NGC~6426 using the ACS photometric catalogue described in the previous section helped us mitigate the broadening of the evolutionary sequences, enabling the derivation of the stellar system age through a one-to-one comparison between the observed CMD and a set of isochrones. The distance modulus, metallicity, and colour excess were also derived simultaneously by 
exploring grids spanning reasonable ranges of parameter values.
To this end, we adopted a Bayesian approach similar to the one used by \citet[][see also \citealp{Saracino2019,Cadelano2019,Cadelano2020b,Deras2023,Deras2024,Cortes2026}]{Cadelano2020a}. To further improve the accuracy of our results, we performed the fit only on stars with a high photometric quality (i.e., sharpness parameter $|sh|\leq0.05$) and located within $200\arcsec$ from the centre reported in the literature (RA~=~17$^{\text h}$~44$^{\text m}$~54.71$^{\text s}$, Dec~=~+03$^{\circ}$~10$^{\arcmin}$~12.51$^{\arcsec}$) \citep{Baumgardt2021}. This selection was also restricted to stars within the magnitude range $17.0~<~\rm m_{\rm F606W}~<~22.0$ to include the Turn-Off, the Sub Giant Branch, and the Red Giant Branch, given that this region is the most sensitive to metallicity and stellar age degeneracies. The isochrones were acquired from the Dartmouth Stellar Evolution Database (DSED) \citep{Dotter2008}\footnote{\url{http://stellar.dartmouth.edu/models/isolf_new.html}}. For each isochrone, we assumed a standard He content $Y$ = 0.25, and $[\alpha \rm /Fe]$ = +0.4, typical values for GCs. We created a grid for each isochrone set, spanning a wide range of ages (10.0 $\leq$ age $\leq$ 20.0 Gyr), metallicities (-2.45 $\leq$ [Fe/H] $\leq$ -1.65), distance moduli (16.0 $\leq$ DM $\leq$ 17.0), and reddening values (0.30 $\leq$ $(E(B-V))$ $\leq$ 0.40). To find the best possible fit, we compared our CMD to the isochrone grid using a Markov Chain Monte Carlo (MCMC) sampling technique. To sample the posterior probability distributions in the $n$-dimensional parameter space, we used the \texttt{emcee} code \citep{Foreman2013,Foreman2019}. Temperature dependent extinction coefficients from \citet{Casagrande2014} were used when converting absolute magnitudes to the observed frame. For the reddening, metallicity, and distance modulus we adopted prior Gaussian distributions centred at $(E(B-V))$ = 0.36 $\pm$ 0.05, [Fe/H] = -2.15 $\pm$ 0.02 \citep{Harris1996}, and DM = 16.58 $\pm$ 0.04 \citep{Baumgardt2021}, respectively, while for the age we assumed a flat prior. Since a flat prior with a hard upper bound can bias the recovered age and its uncertainty when the posterior is not naturally contained within the adopted range, we repeated the fit using three different upper limits on the age prior (13.8, 15.0, and 20.0 Gyr), keeping the extended 10--20 Gyr isochrone grid fixed in all three cases so that the likelihood itself was not artificially restricted by the model grid. The posterior median age shifts systematically with the adopted prior ceiling from 12.7 to 13.3 to 13.9 for the 13.8, 15.0, and 20.0 Gyr priors, respectively. For the two lower ceilings the posterior distribution shows a hard truncation at the prior boundary rather than a natural decline to zero, indicating that the recovered age is prior-dominated rather than data-constrained in those cases. Only when the prior is extended to 20.0 Gyr does the posterior return smoothly to zero on its own, without truncation, showing that this run is no longer limited by the choice of prior. The remaining fitted parameters ($E(B-V)$), true distance modulus and [Fe/H] remain stable across all three runs within their respective uncertainties, indicating that this prior sensitivity is specific to the age determination. We therefore adopt the 20.0 Gyr prior run as our fiducial result and report the age with its natural asymmetric uncertainty, explicitly noting that the data favour an old age for the cluster but do not independently constrain an upper limit. The three posterior distributions are compared in Fig.~\ref{fig:posterior_distribution}, and the corresponding best-fit parameters for each choice of prior are listed in Table~\ref{tab:age_prior_comparison}. It is worth comparing these results with those of \cite{Dotter2011}, who fit DSED isochrones to the same ACS dataset using a fixed reddening and metallicity and a ``by eye'' isochrone fit, for which they report $E(B-V)$=0.36, [Fe/H] = -2.2, DM$_{V}$ = 17.74, and an age of 13.0 $\pm$ 1.5 Gyr. Converting their apparent distance modulus to a true distance modulus using their adopted reddening, $(m-M)_{0}$ = 17.74 - 3.1$\times$0.36 = 16.62, this is consistent with our Bayesian result, $(m-M)_{0}$ = 16.51$_{-0.08}^{+0.08}$, within 1.4$\sigma$. Our reddening ($E(B-V)$ = 0.37$_{+0.02}^{+0.02}$), metallicity ([Fe/H] = -2.15$_{-0.14}^{+0.14}$) and age (13.9$_{-1.7}^{+1.9}$ Gyr) are likewise consistent with theirs within the respective uncertainties. 

\begin{table}[t]
\begin{center}
\caption{Posterior median and the 16th/84th percentile uncertainties for each fitted parameter, as a function of the adopted upper bound (UB) of the flat age prior. The age posterior is prior-dependent, indicating the CMD alone does not independently constrain an upper age limit for NGC~6426.}
\label{tab:age_prior_comparison}
    \begin{tabular}{ccccc} 
    \hline
    \shortstack{Age prior\\UB (Gyr)} & $E(B-V)$ &  $(m-M)_{0}$ & Age (Gyr) & [Fe/H] \\
    \hline
    13.8 & 0.39$_{-0.01}^{+0.01}$ & 16.53$_{-0.07}^{+0.07}$ & 12.7$_{-1.1}^{+0.8}$ & -2.13$_{-0.14}^{+0.14}$ \\
    15.0 & 0.38$_{-0.01}^{+0.02}$ & 16.52$_{-0.07}^{+0.07}$ & 13.3$_{-1.4}^{+1.0}$ & -2.15$_{-0.14}^{+0.14}$ \\
    20.0 & 0.37$_{-0.02}^{+0.02}$ & 16.51$_{-0.08}^{+0.08}$ & 13.9$_{-1.7}^{+1.9}$ & -2.15$_{-0.14}^{+0.14}$ \\
    \hline
\end{tabular}
\end{center}
\end{table}

\begin{figure*}[ht!]
\begin{center}
\includegraphics[scale=0.6]{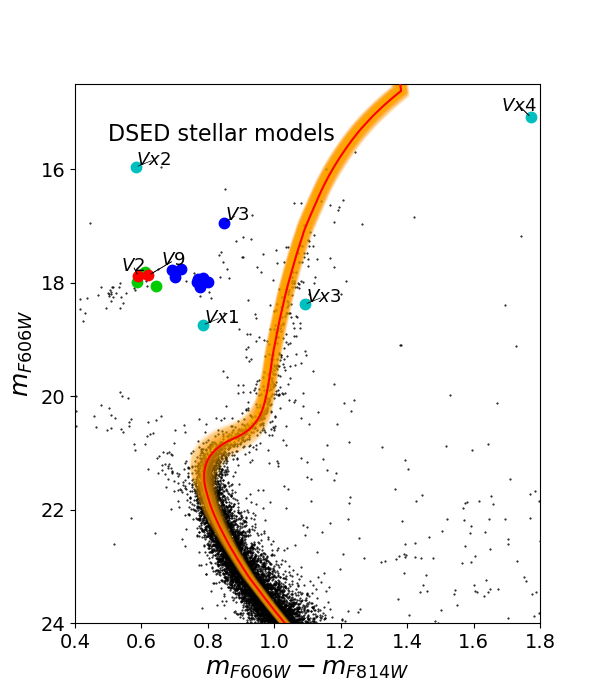}
\includegraphics[scale=0.30]{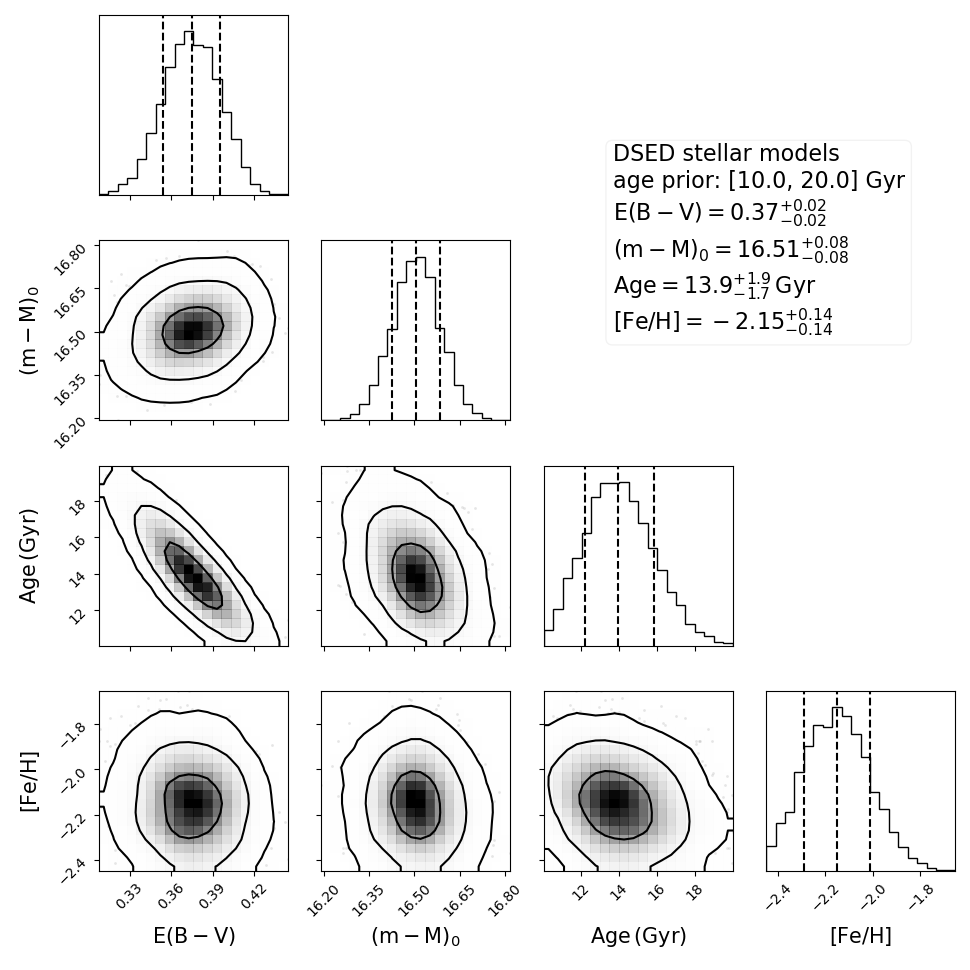}
\caption{{\it Left panel}: CMD of NGC~6426 constructed from the ACS photometry. The best-fit DSED isochrone is plotted as a red solid line and the $1\sigma$-envelope shaded in orange. Blue, green, and red dots correspond to RRab, RRc, and RRd stars, respectively while the turquoise dots correspond to other detected variables.}  {\it Right panel}: Corner plots showing the one- and two-dimensional projections of the posterior probability distributions for all the parameters as obtained from DSED isochrone. The contours correspond to the $1\sigma$, $2\sigma$ and $3\sigma$ levels. \label{fig:isochrones}
\end{center}
\end{figure*}

\begin{figure*}[ht!]
\begin{center}
\includegraphics[scale=0.6]{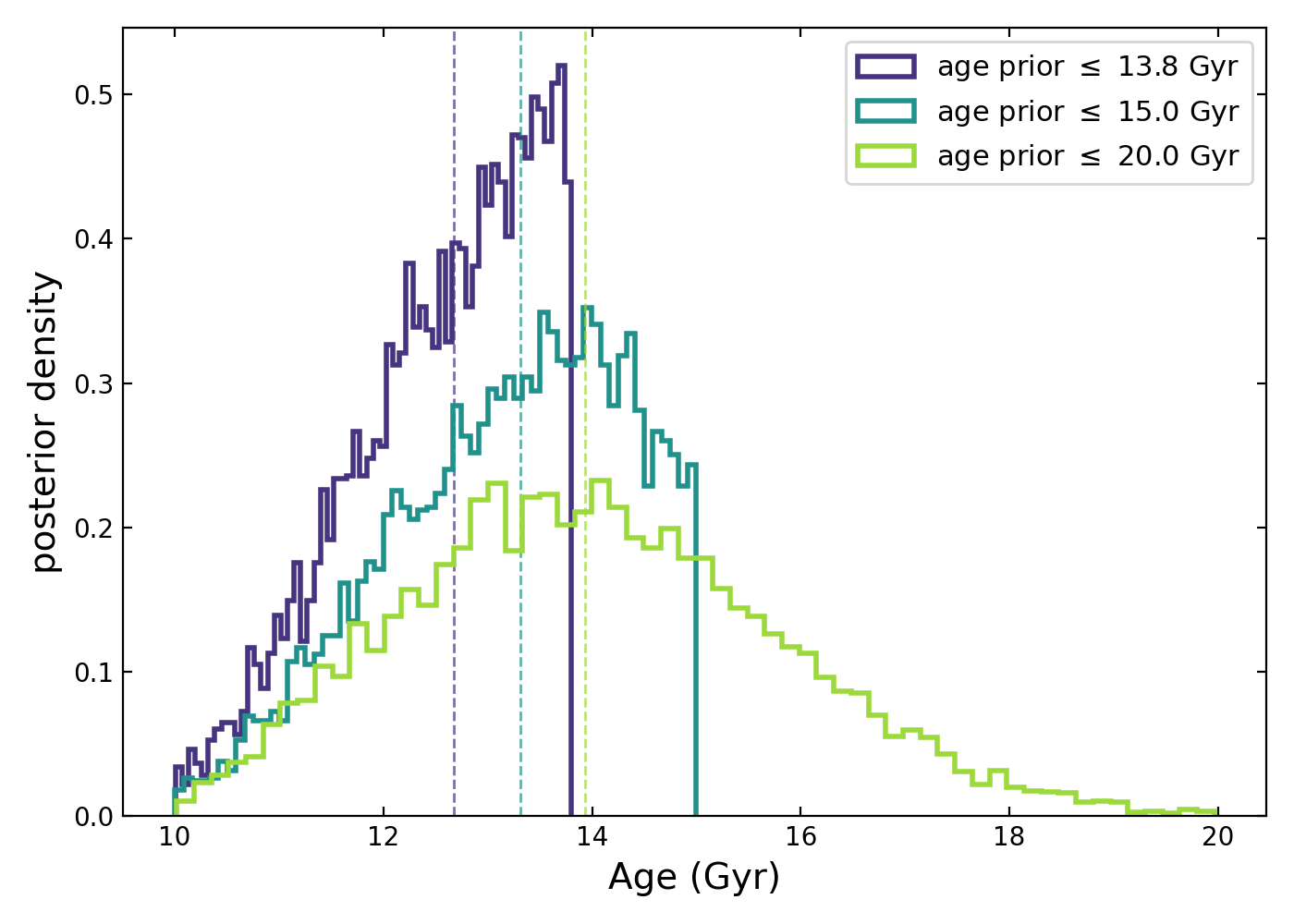}
\caption{Posterior probability distributions of the cluster age derived from three MCMC fits differing only in the upper limit of the flat age prior: 13.8 Gyr (purple), 15.0 Gyr (teal), and 20.0 Gyr (green). All were computed using the same isochrone grid extended to 20.0 Gyr. The dashed vertical lines mark the posterior median age for each run (12.7, 13.3, and 13.9 Gyr, respectively), while solid vertical segments show where the truncated posteriors (13.8 and 15.0 Gyr priors) are cut off by the imposed prior boundary. Only the 20.0 Gyr prior yields a posterior that returns smoothly to zero on its own, showing it is no longer limited by the choice of prior. This run is adopted as the fiducial result.}
\label{fig:posterior_distribution}
\end{center}
\end{figure*}

To place the variable stars on the same CMD as our ACS photometry, we transformed the Johnson-Cousins $VI$ intensity-weighted mean magnitudes reported in Table~\ref{tab:datosgenerales1} into the ACS/WFC $F606W$, $F814W$ system using a theoretical-isochrone-based colour transformation. Briefly, we adopted an isochrone computed for the cluster’s own physical parameters (the one previously fitted with age = 13.9 Gyr and [Fe/H]$~=~-2.15$) which provides synthetic photometry in both photometric systems along a common set of track points. To estimate the offsets $(F606W-V)$ and $(F814W-I)$ at every isochrone point, we expressed them as a function of the isochrones’s intrinsic Johnson colour $(V-I)$, and fitted each relation with a 4th-degree
polynomial. The fits reproduce the isochrone offsets with RMS residuals of 0.002 mag $(F606W-V)$ and 0.0003 mag $(F814W-I)$ over the fitted colour range 0.53 $\leq$ $(V-I)$ $\leq$ 2.00, both well below our typical photometric uncertainties. Finally, we adjusted the position of the variable stars on the CMD by applying the corresponding reddening corrections to their transformed magnitudes (see Fig.~\ref{fig:isochrones}).

\begin{figure*}[ht]
\begin{center}
\includegraphics[width=18cm]{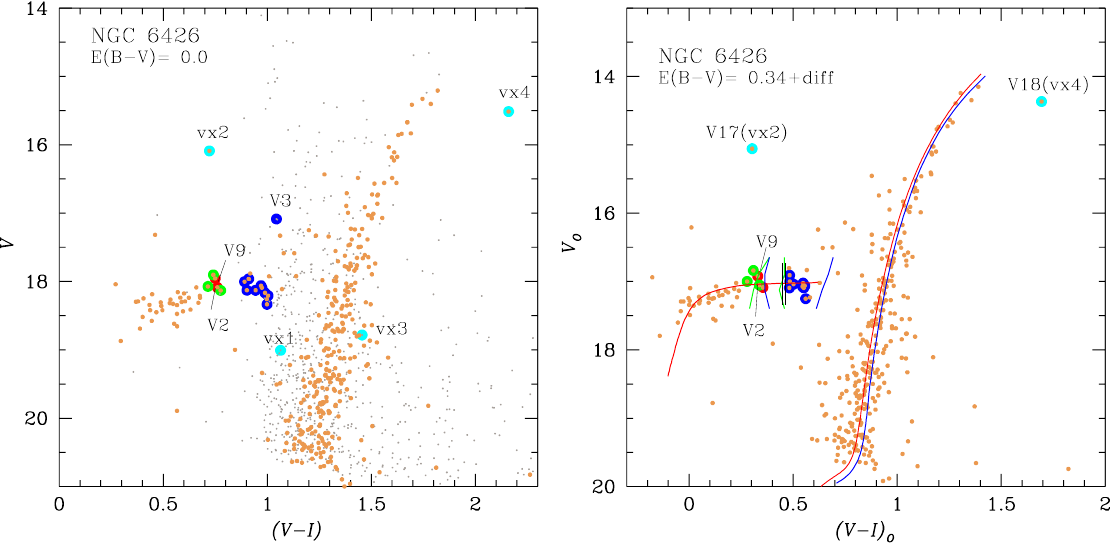}
\caption{\textit{Left panel:} CMD of NGC~6426 constructed from our CCD photometry at SPM. Cluster members are shown in ochre, while field stars are shown in grey. Blue, green and red dots correspond to  RRab, RRc and RRd, respectively, and turquoise indicate other detected variables. Variable stars marked with an ochre dot at their centre are cluster members. The variables vx2 and vx4, both cluster members, are classified as Anomalous  Cepheid and SR star, respectively. We suggest renaming them V17 and V18, respectively to continue with the numbering system adopted by the CVSGC. \textit{Right panel:} Differentially dereddened CMD (see \S~\ref{sec:DifRed}); the RGB and the HB are now more clearly defined.
Field stars have not been plotted in this panel.
The vertical black lines in the HB represent the First Overtone Red Edge (FORE) that separates the fundamental and the first overtone modes; a clean segregation of modes is evident, typical of Oo II type clusters. The continuous red lines represent the isochrone and ZAHB from the Victoria-Regina Models \citep{Vandenberg2014} for [Fe/H] = -2.0, Y= 0.25, [$\alpha$/Fe] = +0.4, and age of 13.5 Gyr. For the sake of comparison, we include in blue colour, the corresponding DSED isochrone fitted in Fig.~\ref{fig:isochrones} but in the $VI$ photometric system.} 

\label{Fig:CMDSPM}
\end{center}
\end{figure*}

\subsection{Dereddening RRab stars}

After the work of \citet{Sturch1966}, it is a well known fact that the $V-I$ colour curves of RRab stars near minimum, between phases 0.5 and 0.8, is constant. This intrinsic colour at minimum was calibrated by \citet{Guldenschuh2005} such that $(V-I)_{0,min} = 0.58 \pm 0.02$. Taking advantage of these results we have calculated the mean $E(V-I)$ in seven of the RRab cluster member stars (top part of Table~\ref{tab:fourier_coeffs}). V7 was omitted due to the large scatter observed in our light curve at these phases. We then calculated $E(B-V) = E(V-I)/1.259$ to obtain the average value $\langle E(B-V)\rangle=0.369 \pm 0.020$. This value can be compared with the predicted by the interstellar reddening calibrations of  \citet{Schlafly2011} $0.298\pm0.012$ and \citet{Schlegel1998} $0.346 \pm 0.014$, at the position of NGC~6426 and with the isochrone fitting result in previous section $E(B-V)$ = $0.37^{+0.02}_{-0.02}$. Due to the presence of differential reddening in the field of the cluster, these small differences are not surprising. 

\subsection{Dereddening CMD and the variable stars}

The CMD constructed from our photometry from SPM is shown in the left panel of Fig.~\ref{Fig:CMDSPM}. A cross-match of the resulting members in \S~\ref{sec:Membership} with our photometric catalogue enabled to distinguish the members stars (ochre dots) from the field stars (small grey dots) in the figure. The positions of all variables discussed in this work are highlighted following the colour code described in the figure caption.

 The CMD in the right panel has been corrected for the differential reddening estimated in \S~\ref{sec:DifRed}. The main sequences in the corrected diagram look better defined. We adopted $E(B-V)=0.34$ to accommodate the variable stars well within the theoretical borders of the instability strip \citep{Bono1995}, and to match the natural mode split displayed by the RRab and RRc distribution with the empirical FORE \citep{Arellano2015,Arellano2016}.

In Fig.~\ref{Fig:CMDSPM} we also include the ZAHB and isochrone for [Fe/H] = -2.0, Y= 0.25, [$\alpha$/Fe] = +0.4, and age of 13.5 Gyr  from the Victoria-Regina Models \citep{Vandenberg2014} (red lines in the figure) and compare it with the version in the $VI$ system of the DSED isochrone fitted on Fig.~\ref{fig:isochrones} to ACS data (blue line).

\section{Summary of results for NGC~6426 and conclusions}
\label{sec:conclusions}
In Table~\ref{summary} we summarise the results for the reddening, metallicity, distance and age of NGC~6426 obtained using an assortment of methods.
For the sake of comparison, included in the bottom part of the table are some values of these parameters found in the literature.

We have peered the membership probabilities of all point sources in the field of our images, which lead to the generation of a CMD free of field stars. Of particular relevance is the membership of the variable stars known in the FoV. Prior to this work, 16 variables were included in the CVSGC. However, V3 has been identified as a field star whereas the membership of V11 (not in our field) and V13 are unknown due to the lack of proper motions data \citep{Prudil2024}.

We confirmed the field status of V3. Our membership analysis  of the four stars independently detected in our photometry, vx1 (RRab), vx2 (AC), vx3 (BY Dra) and vx4 (LPV) (see Appendix~\ref{ap:appendix} for a discussion on their nature) revealed vx2 and vx4 as likely cluster members. We argued in favour of vx2 being a cluster anomalus Cepheid. We suggest to add them to the list of known cluster variables as V17 and V18, to continue with the numbering system in the CVSGC. In summary, the known population of variable stars in NGC~6426 is composed of the 16 stars marked as members (m) in Table~\ref{tab:datosgenerales1}. These are the stars that are considered good indicators of some physical parameters.

The distribution of RRab and RRc stars shows a clear division separated by the empirical FORE instability strip \citep{Arellano2015,Arellano2016} and its theoretical definition \citep{Bono1995}. This is a property of Oo II type clusters \citep{Arellano2015, Yepez2020}.

The usage of theoretical isochrones of  Dartmouth Stellar Evolution Database (DSED), and a deep HST CMD, enable a Bayesian approach to the turn off- giant branch fitting, assuming a Gaussian distribution of priors to  estimate the distance modulus, metallicity, colour excess and cluster age.

NGC~6426 has been associated to the  ancient dwarf galaxy merger event, the Helmi stream \citep{Callingham2022}.
Our results confirm NGC~6426 among the most metal weak and oldest globular clusters in the Galaxy. 

\begin{table*}[t]
\begin{center}
\caption{Summary of results for NGC~6426.}
\label{summary}
    \begin{tabular}{lccccc} 
    \hline
    Method & $E(B-V)$&[Fe/H]$_{\rm ZW}$ &[Fe/H]$_{\rm UVES}$ &  D (kpc) & Age (Gyr) \\
    \hline
& & Present work &  & & \\    
    \hline
    Fourier RRab & &$-1.92 \pm 0.13$ & $-1.97 \pm 0.13$ & $20.96 \pm 0.44$& \\
    Fourier RRc & &$-2.04 \pm 0.17$& $-2.16 \pm 0.17$  & $19.85 \pm 0.70$&   \\
    $I$-band PMZ& &  &  & $21.78 \pm 0.74$& \\
DSED fitting & $0.37^{+0.02}_{-0.02}$&  & $-2.15^{+0.14}_{-0.14}$ & $20.04 \pm 0.70$ & $13.9^{+1.9}_{-1.7}$ \\
 RRab$_{0.5-0.7}$&$0.369 \pm 0.020$ &  &  & & \\

    \hline
Source& & Values in the literature$^{1}$ &  & & \\    
    \hline
 \citet{Schlegel1998} &$0.346 \pm 0.014$&  &  & & \\
\citet{Schlafly2011} & $0.298\pm0.012$&  &  & & \\
\citet{Papadakis2000} & & $-2.16\pm0.13$  &  & & \\

\citet{Harris1996} &$0.36$ & $-2.15$  & &20.6 & \\

Baumgardt et al. 2023$^2$ & &  & &$20.71\pm0.35^3$& \\

\citet{Hatzidimitriou1999} &$0.39\pm0.03$ & $-2.33\pm0.15$  &  &&$\sim 13.45\pm0.25^4$ \\

\citet{SalarisWeiss} & & $-2.20$  & $-2.11$ &&$12.9\pm1.0$ \\

\citet{Dotter2011} && &  & & 13.0$\pm$1.5\\
    \hline
\end{tabular}
\raggedright
\center{
1. Not exhaustive. 2. https://people.smp.uq.edu.au/HolgerBaumgardt/globular/ 3. Adopted value from a set of 8 estimations ranging 19.2-22.0 kpc. 4. Assuming an age for M92 of $12.75\pm 0.25$ Gyr \citep{VandenBerg2013}\\
}
\end{center}
\end{table*}

\section{Acknowlegments}

AAF, RM and MVM are grateful to DGAPA-UNAM for financial support through project  IN103024. We thank the support staff at both the SPM Observatory operated by the Institute of Astronomy, UNAM and the IAO and CREST facilities operated by the Indian Institute of Astrophysics, Bangalore. This work has made use of data from the European Space Agency (ESA) mission Gaia (https://www.cosmos.esa.int/gaia), processed by the Gaia Data Processing and Analysis Consortium (DPAC, https://www.cosmos.esa.int/web/gaia/dpac/consortium). Funding for the DPAC has been provided by national institutions, in particular the institutions participating in the Gaia Multilateral Agreement.


\begin{appendices}
\section{Double mode, new detections and other peculiar stars}
\label{ap:appendix}

{\bf V2}. The light curve amplitude and phase modulations of this star suggest that more than one pulsating mode is active. Subsequent Fourier decompositions enabled the identification of two periods that were in principle identified as the fundamental and first overtone, these are $P_0= 0.543728$ d and $P_1= 0.355447$ d.

\noindent
{\bf V9}. As in the case of V2, the light curve of V9 also suggests a double mode pulsator. The identified periods are $P_0= 0.366649$ d and $P_1=0.289677$ d. 

\begin{figure*}
\begin{center}
\includegraphics[width=17cm,height=9.0cm]{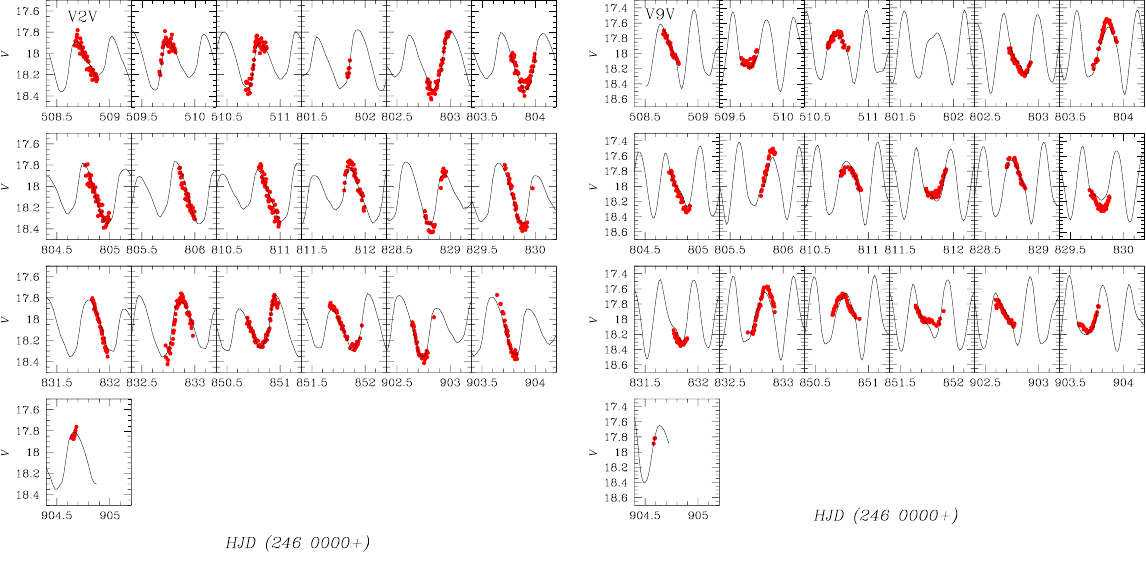}
\caption{V2 and V9 are two double mode stars. The observations (red dots) are fitted with a two-period model (continuous line). See Appendix~\ref{ap:appendix} for a discussion.}
\label{fig:double_mode}
\end{center}
\end{figure*}

In Fig.~\ref{fig:double_mode} we display the observations (red dots) and the resulting model of the two sinusoidals with the corresponding periods. 
The period ratios for V2 and V9 are $P_1/P_0 = 0.65$ and $0.79$, respectively. Both these values are a bit different from the canonical $0.746\pm 0.001$ for radial pulsations \citep{Cox1983},  which may indicate that at least one of the modes is non-radial

\noindent
{\bf vx1}. This star was first reported by \citet{Sesar2017} as an RRab star with a period of 0.6650536174 d and it corresponds to object ZTFJ174454.97+031358.2. The period and shape of its light curve in our data match the original detection. We found the star to be a field star.

\noindent
{\bf vx2} was reported by \citet{Chen2020} as RS CVn with a period of 1.0736 d and corresponds to object ZTFJ174456.30+030847.4. Again, the period and coordinates match our determinations in Table~\ref{tab:datosgenerales1}. This star is a cluster member according to our analysis and in the CMD it lies about two magnitudes above the HB. This would be very rare for a RS CVn star which are normally in the subgiant main sequence. The mean cluster distance and reddening lead to an absolute magnitude $M_V$ of -1.5 to -1.6 mag. For its period of 1.073675 d, the P-L relations for BL Her \citet[][Eq. 3]{Pritzl2003}, yields $M_V\sim$0. If, on the other hand the star is an anomalous Cepheid, its sinusoidal light curve suggests a first-overtone pulsator, for which the P-L relation of \citet[][Eq. 5 and fig. 6]{Pritzl2002} yields $M_V = -1.72 \pm 0.08$ mag, in agreement with the observed value. Thus, in our opinion the star is very likely a cluster member anomalous Cepheid.

\noindent
{\bf vx3} was also reported by \citet{Chen2020} and was classified as a BY Dra variable with a period of 20.8 d, not very different from the determination in our data of 18.08 d. Our light curve is shown in Fig.~\ref{Fig:Vars}. This is a field star not related to the cluster. 

\noindent
{\bf vx4} was studied spectroscopically by \citet{Sharina2012} who classified it as a carbon-rich low-metallicity asymptotic giant branch star. From its radial velocity they conclude it is a cluster member, in agreement with our conclusion based on the $Gaia$-DR3 proper motion analysis. $Gaia$ lists the star as an LPV  (source id: 4376297831805313024) but no periodicity was reported. Our light curve morphology, a period of 39.21 d  (see Table~\ref{tab:datosgenerales1}) and position on the CMD confirm the variability and classification of the star. 

Being vx2 and vx4 authentic cluster member variables, we suggest to include them in the CVSGC as V17 and V18 respectively.
\noindent
{\bf V13}. The cluster membership of this star was not defined in the CVSGC since no $Gaia$-DR3 proper motions are available, thus it was labeled as "unknown". The position of V13 in the log $P$- Amplitude diagram, (Fig.~\ref{Fig:Bailey}) and on the HB in the CMD of Fig.~\ref{Fig:CMDSPM},  strongly argue in favour of its membership to the cluster. It also appears as a likely member in our analysis of the $Gaia$-DR3 proper motions. Therefore, the star was included in the calculation of physical parameters.

\end{appendices}

\renewcommand{\refname}{REFERENCES}
\bibliography{RMxAA.bib}


\end{document}